\documentclass[fleqn,twoside]{article}
\usepackage{espcrc2}

\newcommand{\AmS}{{\protect\the\textfont2
  A\kern-.1667em\lower.5ex\hbox{M}\kern-.125emS}}

\newcommand{\cc}{\cite}

\newcommand{\be}{\begin{equation}}
\newcommand{\ee}{\end{equation}}

\newcommand{\vecc}[1]{\mbox{\boldmath $#1$}}

\def\ex{\hbox{e}}

\def\<{\langle}
\def\>{\rangle}

\def\a{\alpha}

\def\g{\gamma}

\def\d{\delta}
\def\D{\Delta}
\def\l{\lambda}
\def\s{\sigma}
\def\r{\rho}

\def\c{\chi}
\def\m{\mu}
\def\n{\nu}

\def\vf{\varphi}
\def\({\left(}
\def\[{\left[}
\def\){\right)}
\def\]{\right]}
\def\coth{\hbox{coth}}
\def\cot{\hbox{cot}}
\def\cos{\hbox{cos}}
\def\sin{\hbox{sin}}

\def\pa{\mathcal P}

\title{QCD instantons in high energy diffractive scattering: \\ Instanton model of Pomeron\thanks{ Talk given at the International Workshop DIFFRACTION-2004, Cala Gonone, Sardinia, Italy, 18-23 Sept 2004. Supported by RFBR
(04-02-16445, 03-02-17291, 02-02-16194), Russian
Federation President's Grant 1450-2003-2, and the Heisenberg-Landau Program.}}

\author{A.E. Dorokhov\address[MCSD]{Joint Institute for Nuclear Research\\
BLTP JINR, RU-141980 Dubna, Russia}%
        \thanks{Email: {\tt dorokhov@thsun1.jinr.ru}} ,
        I.O. Cherednikov\addressmark[MCSD]\address{Dipartimento di
Fisica, Universit\`a di Cagliari and INFN Sezione di Cagliari
\\ C.P. 170, I-09042  Monserrato (CA), Italy}\thanks{Email: {\tt igorch@thsun1.jinr.ru} }}

\begin{document}

\begin{abstract}
The role of the QCD vacuum effects in high energy diffractive
quark-quark and quark-antiquark scattering is studied with the
Instanton Liquid Model. Special attention is given to the
problem of formation of the soft Pomeron. We show that in the
leading approximation in instanton density  the $C-$odd
instanton contribution to the diffractive amplitude is suppressed
by $1/s$ compared to the $C-$even one. \vspace{1pc}
\end{abstract}

\maketitle

Strong interaction processes in the Regge regime: $s \gg -t $,
{\it i.e.}, collisions with large total center-of-mass energy and
small momentum transfer, yield the main contribution to the
hadronic cross sections at high energy. These processes are
described successfully within the Regge phenomenology, with the
Pomeron exchange being dominant in this regime~\cc{KAI}. The
Pomeron is treated as an effective exchange in the $t$ channel
with vacuum quantum numbers and positive charge parity
$C=+1$. Within perturbative QCD, the pioneer calculations of
the Pomeron properties were performed in Refs.~\cc{BFKL}, which
lead to the supercritical ``hard'' Pomeron violating the Froissart
bound. Further, it was shown that the NLO corrections can change
significantly this leading logarithmic approximation result~\cc{BFKL_NLO}. 
From the other point of view, it is natural to expect that the QCD dynamics 
at large distances and the nontrivial structure of the QCD vacuum 
are relevant for such
processes with small momentum transfer~\cc{LN}. A fruitful
approach is to try to reformulate the complicated QCD dynamics in
terms of some effective theory which would be easier to solve in a given
regime. The convenient effective degrees of freedom at high energy
are the Wilson path-ordered exponentials evaluated along the
straight-line trajectories of colliding particles~\cc{NACH,VERL,LIP,BAL}.

In this work, we consider the high energy diffractive quark-quark
scattering using the Instanton Liquid Model (ILM) of the QCD
ground state~\cc{ILM,DK,REV} in order to take into account
nonperturbative effects in formation of the soft Pomeron. A
similar situation was first analysed in Refs.~\cc{SH,KKLSU} from a
somewhat different point of view.
The quark-quark scattering amplitude is expressed in terms of the
vacuum average of the gauge invariant path-ordered Wilson exponential~\cite{NACH,VERL,KRP} \be T^{kl}_{mn}(s,t) = -2is\int d^2 b_\perp \ex^{i b_\perp q} W^{kl}_{mn}(\c, b^2_\perp) \ ,
\label{Tqq} \ee where the Wilson integral $W^{kl}_{mn}$ reads
\be W^{kl}_{mn}(\c, b_\perp^2)=\Bigg< \pa \ex^{i g \int_{C_{qq}} \! d x_{\m}\hat A_{\m}(x)}
\Bigg\>_0\Bigg|^{kl}_{mn} \ . \label{1ab} \ee In Eq.~(\ref{1ab}),
the corresponding integration path goes along the closed contour
$C_{qq}$: two infinite lines separated by the transverse distance
$\vecc b_\perp$ and having relative scattering angle $\c$. We
parameterize the integration path as $C_\c=\{z_\m(\l);
\l=[-\infty,\infty]\}$ where $z_{\m}(\l)= v_{1}\l \ , \ -\infty < \lambda < \infty$, and
$z_\m(\l) = v_{2}\l + \vecc b \ , \ -\infty < \l < \infty $.
with $\( v_1 v_2\) =\cosh \c$ and $\vecc b = (0_\parallel, b_\perp)$
being the impact parameter in the transversal plane.
To calculate the amplitude (\ref{Tqq}) in the instanton background, we use the explicit expression for
the instanton field \be \hat A_\m (x; \r) = \frac{1}{g_s}
 {\tt R}^{ab} \s^a {\eta^\pm}^b_{\m\n} (x-z_0)_\n \vf
(x-z_0; \r) . \label{if1} \ee The averaging $\< ... \>_0$ over the
nonperturbative vacuum consists in integration over the coordinate
of the instanton center $z_0$, the color orientation ${\tt
R}^{ab}$ and the instanton size $\r$: $dI = d{\tt R} \ d^4 z_0 \
dn_\r $, where the instanton size distribution $dn_\r$ is chosen
according to ILM as~\cc{ILM} $ dn_\r = n_c \d(\r-\r_c) d\r ,$ $
n_c \approx 1 fm^{-4},$ $\r_c \approx 1/3 fm $.
Evaluating the path-ordered exponential Eq.~(\ref{1ab}) and
averaging over all possible embeddings of $SU_c(2)$ into $SU_c(3)$
by using the relations from Ref.~\cc{SVZ80} we get (for further
technical details, see Ref.~\cc{DCH04})
\begin{eqnarray}
& & W_{mn}^{kl}(\g, \vecc b^2) =  n_c \Big\{ \frac{4}{9}
\d_{kl}\d_{mn} \ w_c(\g, \vecc b^2) +
  \nonumber \\
& &  \frac{1}{8} \l^a_{kl}\l^a_{mn} \ \[\frac{1}{3} w_c(\g,\vecc
b^2) + w_s(\g, \vecc b^2)\] \Big\} \  , \label{Tc} \end{eqnarray}
\be w_c (\g, \vecc b^2)= \int \! d^4 z_0 \(\cos \ \a_1 -1 \)
\(\cos \ \a_2 - 1\) \ , \label{wcSc} \ee \be w_s(\g, \vecc b^2) =
- \int \! d^4 z_0 (\hat n^a_1\hat n^a_2)\ \sin \ \a_1 \sin \ \a_2
\ , \ee where the color correlation factor is \be \hat n^a_1\hat
n^a_2 = \frac{(v_1v_2)(z_0,z_0 - \vecc b) - (v_1z_0)(v_2z_0) }
{s_1 s_2 }\ . \ee The phases are defined as \be \a_1 = s_1 \cdot
\int_{-\infty}^\infty \! d\l \ \vf\[( z_0+v_1\l )^2; \r \]\ ,
\label{it22}\ee \be \a_2 =  s_2 \cdot \int_{-\infty}^\infty \! d\l
\ \vf\[(z_0 -v_2 \l - \vecc b)^2; \r \] \ . \ee with $s_1^2 =
z_0^2-(v_1z_0);\ \ \ \ \ s_2^2 = (z_0-\vecc b)^2-(v_2z_0)$. Here
$\g$ is the scattering angle in Euclidean space, whereas in the
final expressions one must make a transition back to Minkowski
space-time (see below). By means of the proper change of variables,
the energy dependence is trivially factorized \be
w_c(\g,b^2_\perp)\to\frac{1}{\sin{\g}}w_c(\pi/2,b^2_\perp) \ ,
\label{fact1} \ee \be w_s(\g,b^2_\perp)\to{\cot\g}\
w_s(\pi/2,b^2_\perp)\ . \label{fact2}\ee

The differential cross section of the quark-quark scattering is
expressed through the amplitude (\ref{Tc}) as \be
\frac{d\sigma_{qq}}{dt}\approx
\frac{1}{9}\frac{1}{s^2}\sum_{kl}\sum_{mn}|T^{kl}_{mn}(s,t)|^2 \ .
\label{dsdt}\ee Making analytical continuation to Minkowski space~\cc{KR,MEG}: 
$ \g \to - i \c $, one finds
$$ \frac{d\sigma_{qq}(t)}{dt} =  \frac{2}{9}n_c^2
\[\coth^2\c\ F_s^2(t) + \right. $$ \be \left. + \frac{2}{3}\frac{\coth\c}{\sinh\c} F_c(t)F_s(t) +
 \frac{11}{3} \frac{1}{\sinh^2\c}F_c^2(t)\] \ ,
\label{dsdt1}\ee where \be F_{s,c}(t) =  \int d^2\vecc b \ex^{i
\vecc b \vecc q} w_{s,c}(\pi/2, \vecc b^2)\ . \ee  In the
asymptotic limit $(\sinh \ \c \sim s \ ,\ \coth \c \to 1)$ the
result (\ref{dsdt1}) coincides with the result of Ref.~\cc{SH}: 
$ \frac{d\s}{dt} \approx \frac{2}{9}n_c^2 F_s^2(t)$ .  In
the weak field limit we reproduce the one-loop single instanton
results (see Ref.~\cc{DCH04}).

For the quark-antiquark scattering
one can treat an antiquark with velocity
$v_2$ as a quark moving backward in time with velocity $-v_2$. As
a result, the scalar product of velocities changes its sign
$(v^q_1v^{\bar q}_2)=-(v^q_1v^q_2)$ and the scattering angles are
related as $ \c_{q q}\to i\pi-\c_{q\bar q}$ .  Then one gets
$$ \frac{d\sigma_{q\bar q}(t)}{dt} =  \frac{2}{9}n_c^2
\[\coth^2\c\ F_s^2(t) - \right. $$ \be \left.  - \frac{2}{3}\frac{\coth\c}
{\sinh\c}F_c(t)F_s(t) + \frac{11}{3}\frac{1}{\sinh^2\c}F_c^2(t)\]
\ . \label{dsdt1qbq} \ee The second terms in Eqs.~(\ref{dsdt1}) and
(\ref{dsdt1qbq}) correspond to the contribution of the $C-$odd
amplitude.

The spin averaged total quark-quark cross section in the
instanton--antiinstanton approximation reads $$
\s_{qq} \approx $$ \be \frac{2}{9}n_c^2\int_0^\infty d\vecc q^2
\[F_s^2(\vecc q^2)
+ \frac{4}{3}\frac{m^2}{s}F_c(\vecc q^2)F_s(\vecc q^2)\]
\label{Sas}\ee which is constant in the high energy limit. It is
finite if the constrained instanton solution is used~\cc{DEMM99}.
In Eq.~(\ref{Sas}), the only term corresponding to the $C=+1$
exchange, Pomeron, survives in the asymptotics, while the $C=-1$
contribution (second term in Eq.~(\ref{Sas})) is
suppressed by the small factor $\sim m^2/s$. The leading $C=-1$
part of the scattering amplitude, odderon, will arise at higher orders
in instanton density, which corresponds to diagrams like
three nonperturbative gluon exchange. The growing part of
the total cross section $ \D\s _{qq} \sim (n_{c}\rho _{c}^{4})^2\D
(t)\ln s $ can arise only if {\it inelastic} quark-quark
scattering in the instanton-antiinstanton background is taken into
account~\cite{SH}.

It is important to note that the original Wilson exponential,
Eq.~(\ref{1ab}), has essentially Minkowskian light-cone geometry
whereas the instanton calculations of Wilson loop are performed in
the Euclidean QCD. The mapping from the Minkowski space to the Euclidean
one is possible since the dependence of the Wilson integral on the
total energy $s$ and transverse momentum squared $t$ is factorized
in Eqs.~(\ref{fact1}), (\ref{fact2}). At high energy, the
amplitude is $s-$independent both in the perturbative and
nonperturbative cases. At the same time, the $t-$dependence of the
amplitude is naturally expressed through the nonperturbative
instanton field.

\noindent {\small The authors thank the Organizers of
Diffraction'04 for warm hospitality and financial support during
the Workshop. I.Ch. thanks also ICTP (Trieste) for invitation and
support.}

\end{document}